\documentclass[conference]{IEEEtran}
\IEEEoverridecommandlockouts
\usepackage{cite}
\usepackage{amsmath,amssymb,amsfonts}
\usepackage{algorithmic}
\usepackage{graphicx}
\usepackage{textcomp}
\usepackage{xcolor}
\usepackage{url} 
\usepackage{array}      
\usepackage{tabularx}   
\def\BibTeX{{\rm B\kern-.05em{\sc i\kern-.025em b}\kern-.08em
    T\kern-.1667em\lower.7ex\hbox{E}\kern-.125emX}}
\begin{document}

\title{L2T-Tune:LLM‑Guided Hybrid Database Tuning with LHS and TD3\\

\thanks{*Chen Zheng is the corresponding author.
}
}

\author{\IEEEauthorblockN{1\textsuperscript{st} Xinyue Yang}
\IEEEauthorblockA{\textit{University of Chinese Academy of Sciences, Nanjing} \\
\textit{Institute of Software, Chinese Academy of Sciences}\\
Nanjing, China \\
yangxinyue241@mails.ucas.ac.cn}
\\
\IEEEauthorblockN{3\textsuperscript{rd} Yaoyang Hou}
\IEEEauthorblockA{\textit{Hangzhou Institute for Advanced Study, UCAS} \\
Hangzhou, China \\
houyaoyang23@mails.ucas.ac.cn}

\\
\IEEEauthorblockN{5\textsuperscript{th} Yinyan Zhang}
\IEEEauthorblockA{\textit{University of Chinese Academy of Sciences, Nanjing} \\
\textit{Institute of Software, Chinese Academy of Sciences}\\
Nanjing, China \\
zhangyinyan23@mails.ucas.ac.cn}
\and
\IEEEauthorblockN{2\textsuperscript{nd} Chen Zheng\textsuperscript{*}}
\IEEEauthorblockA{\textit{Institute of Software, Chinese Academy of Sciences} \\
\textit{University of Chinese Academy of Sciences, Nanjing}\\
\textit{Hangzhou Institute for Advanced Study, UCAS}\\
Beijing, China \\
zhengchen@iscas.ac.cn}
\\

\IEEEauthorblockN{4\textsuperscript{th} Renhao Zhang}
\IEEEauthorblockA{\textit{Hangzhou Institute for Advanced Study, UCAS} \\
Hangzhou, China \\
zhangrenhao23@mails.ucas.ac.cn}
\\
\IEEEauthorblockN{6\textsuperscript{th} Yanjun Wu}
\IEEEauthorblockA{\textit{Institute of Software, Chinese Academy of Sciences} \\
Beijing, China \\
yanjun@iscas.ac.cn}

\\
\IEEEauthorblockN{7\textsuperscript{th} Heng Zhang}
\IEEEauthorblockA{\textit{Institute of Software, Chinese Academy of Sciences} \\
Beijing, China \\
zhangheng17@iscas.ac.cn}
}

\maketitle

\begin{abstract}
Configuration tuning is critical for database performance. Although recent advancements in database tuning have shown promising results in throughput and latency improvement, challenges remain. First, the vast knob space makes direct optimization unstable and slow to converge. Second, reinforcement learning pipelines often lack effective warm-start guidance and require long offline training. Third, transferability is limited: when hardware or workloads change, existing models typically require substantial retraining to recover performance.

To address these limitations, we propose L2T-Tune, a new LLM-guided hybrid database tuning framework that features a three-stage pipeline: Stage one performs a warm start that simultaneously generates uniform samples across the knob space and logs them into a shared pool; Stage two leverages a large language model to mine and prioritize tuning hints from manuals and community documents for rapid convergence. Stage three uses the warm-start sample pool to reduce the dimensionality of knobs and state features, then fine-tunes the configuration with the Twin Delayed Deep Deterministic Policy Gradient algorithm.

We conduct experiments on L2T-Tune and the state-of-the-art models. Compared with the best-performing alternative, our approach improves performance by an average of 37.1\% across all workloads, and by up to 73\% on TPC-C. Compared with models trained with reinforcement learning, it achieves rapid convergence in the offline tuning stage on a single server. Moreover, during the online tuning stage, it only takes 30 steps to achieve best results.
\end{abstract}


\section{Introduction}
Modern database management systems (DBMSs) serve as the backbone of data-intensive applications, ranging from e-commerce platforms to financial transaction systems.
Performance, dictated by throughput, latency, and resource efficiency, is heavily influenced by numerous configuration parameters, known as knobs.
Modern database systems like MySQL and PostgreSQL offer hundreds of tunable knobs for memory allocation, query execution, logging, and concurrency control.
The inter-dependencies among these knobs make manual tuning by database administrators (DBAs) laborious and prone to suboptimal outcomes, especially as workload characteristics and hardware specifications change \cite{CDBTune,duan2009tuning}.

Recently, recognizing the impracticality of manual tuning at scale, the database research community has pursued automated approaches that span multiple paradigms. Various methods have been proposed for automating database tuning.
Search-based tuning, as shown by BestConfig \cite{BestConfig}, partitions the knob space and performs extensive exploration, which reduces manual effort, but rarely finds truly optimal settings in a limited time. Because it must probe large portions of the space, a new tuning request or environment change typically requires restarting the process, offering limited adaptability.

Researchers use machine learning for database tuning \cite{van2017automatic,gallinucci2019sparktune,cgptuner,duan2009tuning,zhang2021restune,zhang2022towards}. OtterTune \cite{van2017automatic} uses Gaussian process regression and internal metrics instead of just TPS/latency/QPS, offering better guidance. However, it tunes only about ten parameters on PostgreSQL, reducing its effectiveness in high-dimensional spaces.

Reinforcement learning (RL) approaches\cite{HUNTER,li2019qtune,CDBTune}, exemplified by CDBTune \cite{CDBTune} and HUNTER \cite{HUNTER}, use actor–critic updates such as DDPG \cite{lillicrap2015continuous}, Q-learning \cite{watkins1992q} and DDQN \cite{van2016deep}. Compared with traditional machine learning, they need less precollected data and thus reduce measurement overhead. CDBTune combines offline training with online fine-tuning, which provides good transferability across environments. The study "Too Many Knobs to Tune?" \cite{kanellis2020too} shows that adjusting only a few knobs can already deliver substantial gains, which motivates HUNTER to reduce the dimensionality of both the knobs and the state of the system. Importantly, when hardware or workloads change, CDBTune and HUNTER do not need to be retrained from scratch; their policies can be adapted online, yet reaching a new optimum typically still takes several hours of interaction. In addition, early RL exploration remains weakly guided with poor warm starts, and many systems depend on large-scale parallel servers, leading to significant resource costs.

With the rise of large language models\cite{DB-BERT,LLMTune,GPTuner,lambdatune,Latuner,huang2024e2etune}, systems such as DB-BERT \cite{DB-BERT}, LLMTune \cite{LLMTune} and GPTuner \cite{GPTuner} mine posts and official manuals, preprocess the text, and prompt an LLM to recommend knob settings, much like consulting an experienced DBA for rapid, workload-aware adjustments. These methods avoid exhaustive space search and enable fast online recommendations. However, because the tuned knobs are drawn from documentation, the actionable set is narrow and biased toward common parameters. While they converge quickly to a reasonable baseline, they often remain measurably short of the true optimum.

To address these challenges, we propose L2T-Tune, a Three-Stage LLM-Guided Hybrid Database Tuning with Latin hypercube sampling (LHS) \cite{mckay2000comparison} and Delayed Deep Deterministic Policy Gradient algorithm (TD3) \cite{fujimoto2018addressing}. Our model follows the offline training + online fine-tuning framework adopted from CDBTune. The three stages of our model are as follows.

\begin{itemize}
    \item \textbf{Stage One}: LHS warm start. We use LHS to initialize with uniformly distributed configurations, yielding a stronger starting point and a well-spread dataset for later dimensionality reduction.

    \item \textbf{Stage Two}: LLM-guided recommendation. We employ DB-BERT and the optimized variant of GPTuner, two state-of-the-art LLM-based tuners, to achieve rapid convergence. DB-BERT provides documentation-derived hints, while GPTuner provides structured range suggestions. In our setting, the optimized GPTuner runs coarse-only tuning in the tiny feasible space and omits the fine stage. Their outputs yield a baseline for subsequent fine-tuning.

    \item \textbf{Stage Three}: Random Forest (RF)\cite{RF}/Principal Components Analysis (PCA)\cite{PCA} + TD3 fine-tuning. Using the Stage-One samples, RF selects impactful knobs and PCA compresses state features. We then perform full TD3 reinforcement learning fine-tuning to reach the final optimum.

\end{itemize}

During the reinforcement learning tuning process, 63 parameters are used to represent the current state of MySQL, which guides the actor's actions in the TD3 model. These 63 parameters are reduced in dimensionality using PCA, further enhancing efficiency. Finally, the TD3 model is used for reinforcement learning training. The TD3 model continuously updates the actor and critic networks based on feedback from the environment, fine-tuning the database knobs for optimal performance.

Additionally, in online fine-tuning we use a semi-transfer scheme: when memory or disk changes, we migrate the Stage-1 LHS warm start (sample pool and knob bounds), apply the LLM model for rapid re-baselining, and then run a brief TD3 local fine-tune. This quickly absorbs hardware shifts and improves transfer performance with minimal online steps.

The key contributions of this work are:

\begin{enumerate}
    \item \textbf{Ensuring Uniform Data Sampling in the Warm Start:} By utilizing LHS for the warm start, our model not only improves initial performance, but also ensures more uniform data sampling. This uniform sampling provides a better foundation for the subsequent stages of dimensionality reduction, making the entire tuning process more efficient.

\item \textbf{Combining RF with LLMs:} Our approach combines RL with LLMs to create a hybrid offline training and online fine-tuning framework. This integration leverages the adaptability and transferability of RL for continuous learning, while benefiting from the fast convergence of LLMs, resulting in a more efficient and quicker tuning process.

\item \textbf{Rapid Offline Convergence on a Single Server:}During the offline training stage, our model converges rapidly and reaches the optimal configuration on a single server. Compared with Hunter and CDBTune, it does not require a large amount of parallel computing resources.

\item \textbf{Improving Transferability:} In the online fine-tuning phase, we enhance transferability by incorporating rapid recommendations from the large language model. Compared to traditional reinforcement learning models, this approach significantly improves transfer performance while maintaining a short training time.

\item \textbf{Superior Tuning Performance:} Across workloads, our approach achieves an average improvement of 37.1\%, reaches up to 73\% on TPC-C, and requires only about 30 online-tuning steps to reach the best configuration.

\end{enumerate}

\section{Preliminary}
To maintain consistency with the offline training + online fine-tuning framework used in CDBTune and HUNTER, we adopt the same data format representation. Additionally, we implement a shared pool mechanism to efficiently store and manage the data samples generated during the tuning process. Each tuning iteration represents a sample as a tuple 
 \((S, A, P)\), where:

 \paragraph{\textbf{A (Action/Configuration)}} A represents the configuration of MySQL knobs---the adjustable parameters that control database behavior. Modern MySQL exposes 266 tunable knobs governing memory allocation (e.g., \url{innodb_buffer_pool_size}), concurrency control (e.g., \url{max_connections}), I/O behavior (e.g., \url{innodb_flush_log_at_trx_commit}), query optimization (e.g., \url{optimizer_search_depth}), and logging (e.g., \url{innodb_log_file_size}). Similarly, PostgreSQL typically offers 346 tunable knobs, controlling similar categories of database performance and behavior.

  \paragraph{\textbf{P (Performance Metrics)}} The external performance under configuration \(A\) measured via benchmarks: (1) Throughput (TPS) (2) p95 Latency (3) QPS. These serve as the optimization objective.

  \paragraph{\textbf{S (State/Internal Metrics)}} To avoid sparse rewards, we augment the state with 63 internal metrics from \url{performance_schema} and \url{information_schema.INNODB_METRICS}, enabling the agent to relate configuration changes to behaviors (e.g., increased buffer pool misses \(\rightarrow\) lower TPS). L2T-Tune adopts this 63-dimensional representation.

The 63 internal metrics are organized into five categories, as detailed in Table~\ref{tab:innodb-metrics}.

\begin{table*}[t]
\caption{The 63 Internal Metrics from MySQL \texttt{INNODB\_METRICS}}
\label{tab:innodb-metrics}
\centering
\setlength{\tabcolsep}{3pt}
\renewcommand{\arraystretch}{1.1}
{\scriptsize
\begin{tabularx}{\textwidth}{>{\raggedright\arraybackslash}p{2.3cm}>{\centering\arraybackslash}p{0.7cm}>{\raggedright\arraybackslash}X>{\raggedright\arraybackslash}p{3.2cm}}
\hline
\textbf{Category} & \textbf{Count} & \textbf{Metrics} & \textbf{Description}\\
\hline

\textbf{State Metrics} & (14) &
\texttt{metadata\_mem\_pool\_size}, \texttt{lock\_row\_lock\_time\_max}, \texttt{lock\_row\_lock\_time\_avg}, \texttt{buffer\_pool\_size}, \texttt{buffer\_pool\_pages\_total}, \texttt{buffer\_pool\_pages\_misc}, \texttt{buffer\_pool\_pages\_data}, \texttt{buffer\_pool\_bytes\_data}, \texttt{buffer\_pool\_pages\_dirty}, \texttt{buffer\_pool\_bytes\_dirty}, \texttt{buffer\_pool\_pages\_free}, \texttt{trx\_rseg\_history\_len}, \texttt{file\_num\_open\_files}, \texttt{innodb\_page\_size}
& Instantaneous values reflecting current state (e.g., buffer pool size, dirty pages, open files).\\[2pt]

\textbf{Current/Instant Metrics} & (13) &
\texttt{lock\_row\_lock\_current\_waits}, \texttt{buffer\_pool\_read\_ahead\_evicted}, \texttt{ibuf\_merges\_discard\_delete\_mark}, \texttt{innodb\_rwlock\_s\_spin\_rounds}, \texttt{innodb\_rwlock\_x\_spin\_rounds}, \texttt{innodb\_rwlock\_s\_os\_waits}, \texttt{innodb\_rwlock\_x\_os\_waits}, \texttt{innodb\_dblwr\_pages\_written}, \texttt{innodb\_rwlock\_s\_spin\_waits}, \texttt{innodb\_rwlock\_x\_spin\_waits}, \texttt{ibuf\_merges\_discard\_delete}, \texttt{buffer\_pool\_read\_requests}, \texttt{buffer\_pool\_write\_requests}
& Instantaneous counters of ongoing operations (e.g., lock waits, read-ahead activity, spin waits).\\[2pt]

\textbf{Cumulative-1} & (12) &
\texttt{lock\_row\_lock\_time}, \texttt{lock\_row\_lock\_waits}, \texttt{buffer\_pool\_wait\_free}, \texttt{buffer\_pool\_read\_ahead}, \texttt{adaptive\_hash\_searches}, \texttt{adaptive\_hash\_searches\_btree}, \texttt{ibuf\_merges\_delete\_mark}, \texttt{ibuf\_merges\_discard\_insert}, \texttt{os\_log\_pending\_fsyncs}, \texttt{os\_log\_pending\_writes}, \texttt{os\_log\_bytes\_written}, \texttt{innodb\_activity\_count}
& Cumulative (frame-differenced) counters for lock contention, buffer pool waits, AHI usage, and log I/O.\\[2pt]

\textbf{Cumulative-2} & (12) &
\texttt{buffer\_pages\_written}, \texttt{buffer\_pages\_read}, \texttt{buffer\_data\_reads}, \texttt{buffer\_data\_written}, \texttt{ibuf\_merges\_insert}, \texttt{ibuf\_merges\_delete}, \texttt{innodb\_dblwr\_writes}, \texttt{buffer\_pool\_reads}, \texttt{buffer\_pages\_created}, \texttt{log\_write\_requests}, \texttt{os\_data\_reads}, \texttt{os\_data\_writes}
& Cumulative counts for buffer-pool I/O, insert buffer activity, doublewrite buffer, and OS-level data ops.\\[2pt]

\textbf{Cumulative-3} & (12) &
\texttt{os\_data\_fsyncs}, \texttt{os\_log\_fsyncs}, \texttt{lock\_deadlocks}, \texttt{lock\_timeouts}, \texttt{log\_waits}, \texttt{log\_writes}, \texttt{ibuf\_merges}, \texttt{ibuf\_size}, \texttt{dml\_reads}, \texttt{dml\_inserts}, \texttt{dml\_deletes}, \texttt{dml\_updates}
& Cumulative counters for fsyncs, lock failures, log writes, insert buffer state, and DML operations.\\
\hline
\end{tabularx}
}
\end{table*}

These metrics are collected through a multi-frame sampling strategy to reduce noise: during each benchmark execution (e.g., Sysbench runs for 60\,s with \url{--time=60}), a background timer collects internal metrics from \url{INNODB_METRICS} every 5\,s, yielding \(T\) frames that align with the benchmark duration (\(T=12\) for 60\,s runs). Let \(m_i^t\) denote the value of metric \(m_i\) at frame \(t \in \{1,2,\ldots,T\}\). We aggregate these temporal observations into a single scalar \(s_i\) for each metric using two strategies:

\textbf{Type 1: Cumulative Counters} (e.g., \url{lock_row_lock_time}, \url{os_data_fsyncs}, \url{buffer_pages_written}). These monotonically increasing counters track cumulative events since MySQL startup. To capture the \emph{rate} of activity during the observation window, we compute the difference between the last and first frames.

\begin{equation}
s_i^{\text{counter}} = m_i^T - m_i^1 .
\end{equation}

\textbf{Type 2: Instantaneous Values} (e.g., \url{buffer_pool_pages_dirty}, \url{lock_row_lock_current_waits}, \url{buffer_pool_size}). These metrics represent the current state at each sampling instant and may fluctuate due to transient workload spikes or background processes. To obtain a stable representative value, we compute the temporal average:
\begin{equation}
s_i^{\text{instant}} = \frac{1}{T} \sum_{t=1}^{T} m_i^t .
\end{equation}

This averaging smooths out short-term variations, providing a robust signal that captures the typical operating regime during the trial. The resulting 63-dimensional state vector \(\mathbf{S} = [s_1, s_2, \ldots, s_{63}]\) is then fed to the RL agent (after optional PCA compression in Stage~3) to guide policy learning.

\section{L2T-Tune: System Architecture and Workflow}
\begin{figure*}[t]
  \centering
  \includegraphics[width=\textwidth]{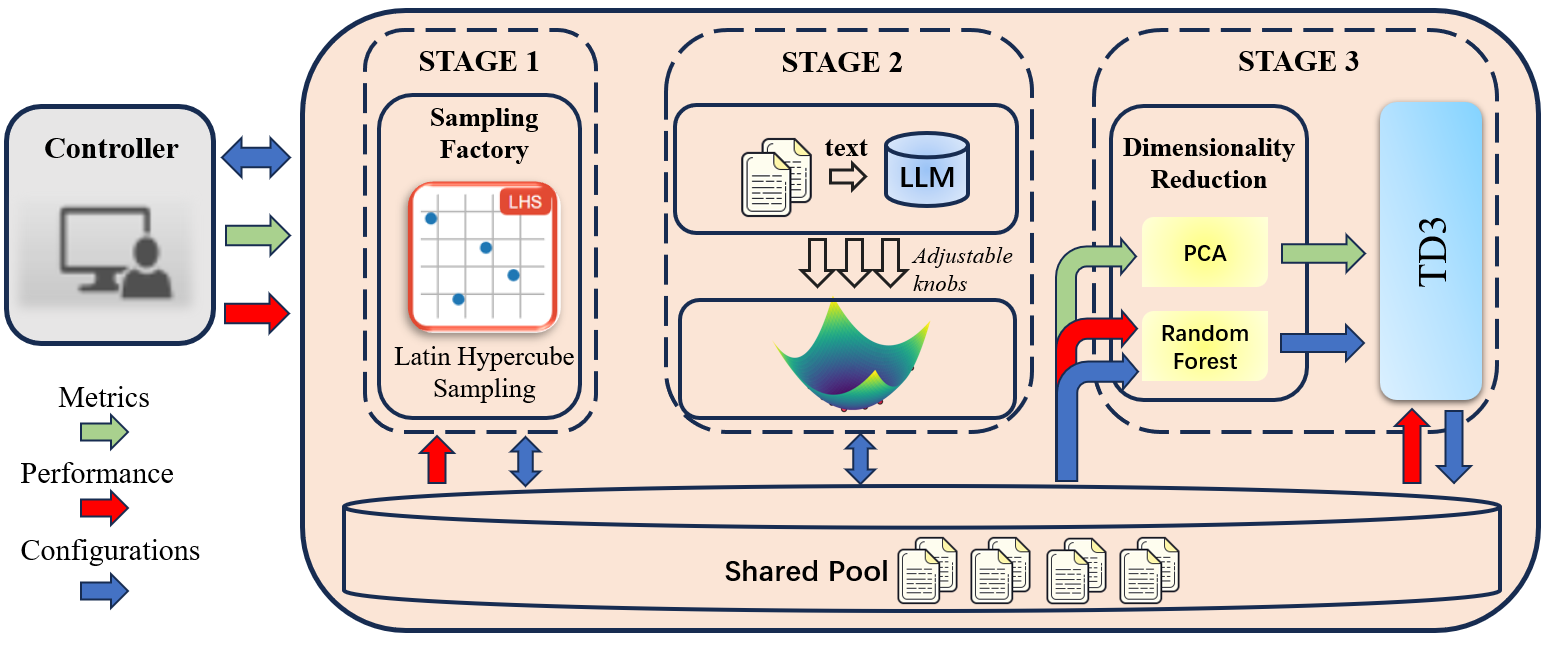} 
  \caption{Architecture of L2T-Tune.}
  \label{fig:overview}
\end{figure*}

This section presents the architecture of L2T-Tune, an online hybrid database tuning system designed to address the practical challenges of cloud database configuration optimization. L2T-Tune builds upon and extends the Hunter framework with a novel three-stage pipeline that synergistically combines sampling-based exploration, LLM-guided semantic reasoning, and reinforcement learning-based tuning to achieve faster convergence and superior performance. Figure~\ref{fig:overview} illustrates the overall data flow and integration across stages. During the entire tuning process, the data is stored in a shared pool as samples represented by the tuple \((S, A, P)\), where each sample consists of the state (S), the action (A), and the resulting performance (P).

\subsection{Stage 1: LHS Warm-Start for Diverse Exploration}

CDBTune applies DDPG to tune knobs from scratch, beginning with random exploration. This cold start faces two hurdles: (1) the high-dimensional knob space makes unguided exploration inefficient, often requiring thousands of trials; (2) early rewards are sparse/noisy, slowing convergence. HUNTER mitigates these issues with a GA-based warm start that seeds the replay buffer with diverse, higher-quality candidates, and further accelerates wall-clock time by using cloned database instances to evaluate multiple configurations in parallel on identical snapshots. This combination speeds early learning and reduces variance across trials.

We adopt LHS~\cite{mckay2000comparison} for warm start instead of GA. ITuned first brought LHS to database tuning and showed that its space-filling property enables efficient exploration of high-dimensional configuration spaces with limited budgets, outperforming random and grid sampling. LHS offers the advantage of uniform sampling, ensuring that the parameter space is evenly covered, which helps avoid clustering often seen with the stochastic operators of genetic algorithms.

Take MySQL as an example, to generate the LHS samples, we initialize an LHS sampler with \(d=266\) and generate \(n=120\) normalized action vectors \(\{\mathbf{a}^{(1)}, \ldots, \mathbf{a}^{(120)}\} \subset [0,1]^{266}\).

In each iteration, we map the action vector \(\mathbf{a}^{(j)}\) to the physical knobs, applying trust-region constraints with a \url{trust_ratio} of \(0.05\). We then apply the configuration and run the load benchmark. Finally, we collect the resulting metrics and record the sample \(\big(S^{(j)}, A^{(j)}, P^{(j)}\big)\) into the shared pool for further analysis.

\begin{figure}[t]
  \centering
  \includegraphics[width=\columnwidth]{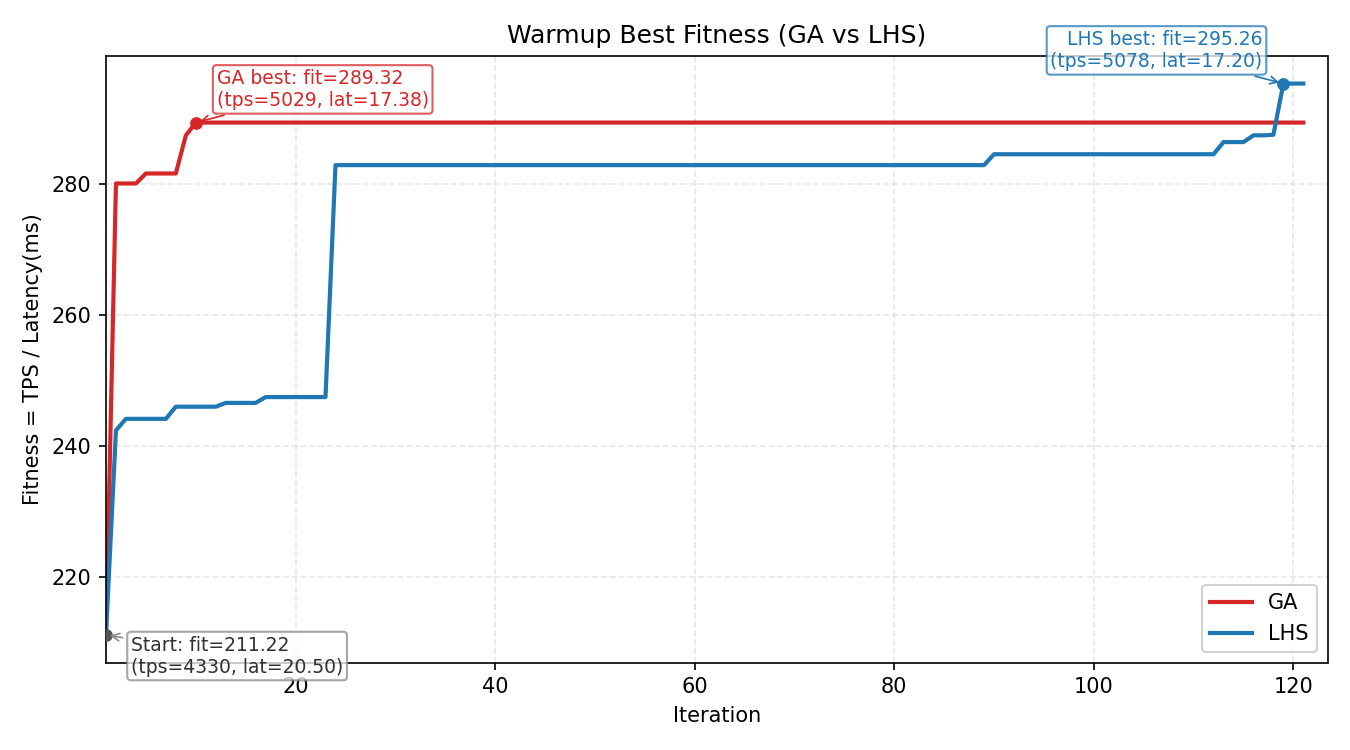} 
  \caption{Best fitness evolution for GA vs.\ LHS warm start (120 iterations, Sysbench read).}
  \label{fig:gavslhs}
\end{figure}

To empirically validate our choice of LHS over GA, we conducted a controlled comparison on the Sysbench read workload. Both methods executed 120 tuning iterations under identical conditions:

Figure~\ref{fig:gavslhs} shows the best-fitness curve over 120 iterations, where fitness is defined as
\begin{equation}
f \;=\; \frac{\mathrm{TPS}}{\text{p95 latency}}.
\end{equation}

LHS achieves a best fitness of \textbf{295.26} (TPS \(=5078\), latency \(=17.20\)\,ms), compared to GA's \textbf{289.32} (TPS \(=5029\), latency \(=17.38\)\,ms)---a \textbf{2.1\%} improvement. The effect of LHS is slightly better than that of GA, and this has also been verified through actual testing in other loads.

Beyond improving warm-start configurations, the stratified sampling of LHS ensures more uniform coverage across all knobs. This uniformity serves two key purposes: (1) it raises the performance baseline for subsequent LLMs (Stage~2) and TD3 (Stage~3), minimizing the impact of poor initializations, and (2) it creates a well-distributed dataset crucial for Random-Forest-based knob selection in Stage~3, allowing for robust feature-importance estimation without the bias toward clustered regions that GA tends to introduce. By combining enhanced warm-start performance with consistent data sampling for dimensionality reduction, LHS lays a strong foundation for L2T-Tune’s three-stage pipeline.

\subsection{Stage 2: LLM-guided recommendation.}

Recent LLM-based tuners can read vendor manuals, blogs, and forums to synthesize knob recommendations, achieving fast convergence with only a few dozen trials. Two representative systems are GPTuner and DB-BERT. Because they extract different forms of evidence—structured ranges vs.\ free-text hints—we treat Stage~2 as a pluggable module and instantiate it with either backbone.

\subsubsection{Improved GPTuner}

\begin{figure}[t]
  \centering
  \includegraphics[width=\columnwidth]{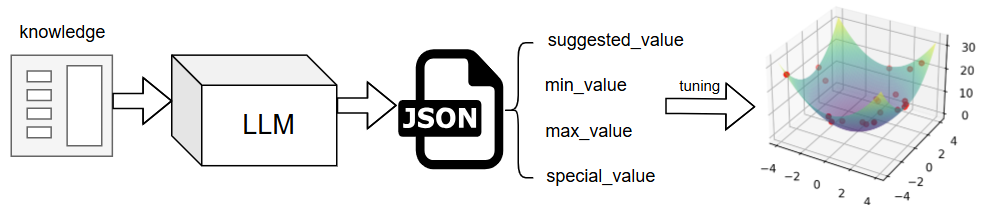} 
  \caption{Overview of GPTuner.}
  \label{fig:gptuner}
\end{figure}

\textbf{Sources \& data schema.}
GPTuner ``reads the manual'' by harvesting domain knowledge from official documentation, blogs/forums, and GPT-generated text. The pipeline cleans and integrates these texts into a machine-readable \emph{structured} form per knob (JSON with \url{suggested_values}, \url{min_value}, \url{max_value}, and \url{special_value}), and organizes them in a Tuning Lake. This lets downstream optimizers reason over realistic ranges instead of vendor-wide defaults.
As illustrated in Figure~\ref{fig:gptuner}, the overall pipeline consists of a Knowledge Handler, a Search-Space Optimizer, and a Configuration Recommender.

\textbf{Two-stage tuning.}
GPTuner then uses a coarse-to-fine search. In the coarse stage, it builds a Tiny Feasible Space from documentation suggestions and range-scaled variants, seeds it with a small LHS design, fits a random-forest surrogate, and runs SMAC \cite{lindauer2022smac3} (RF-based Bayesian optimization) for a few iterations. The fine stage then widens the search to the full heterogeneous ranges and continues SMAC, bootstrapped by the coarse samples, to refine numeric values. The pipeline has three parts: a Knowledge Handler that builds the Tuning Lake, a Search-Space Optimizer for knob selection and range shaping, and a Configuration Recommender that executes the coarse-to-fine loop.

\begin{figure}[t]
  \centering
  \includegraphics[width=\columnwidth]{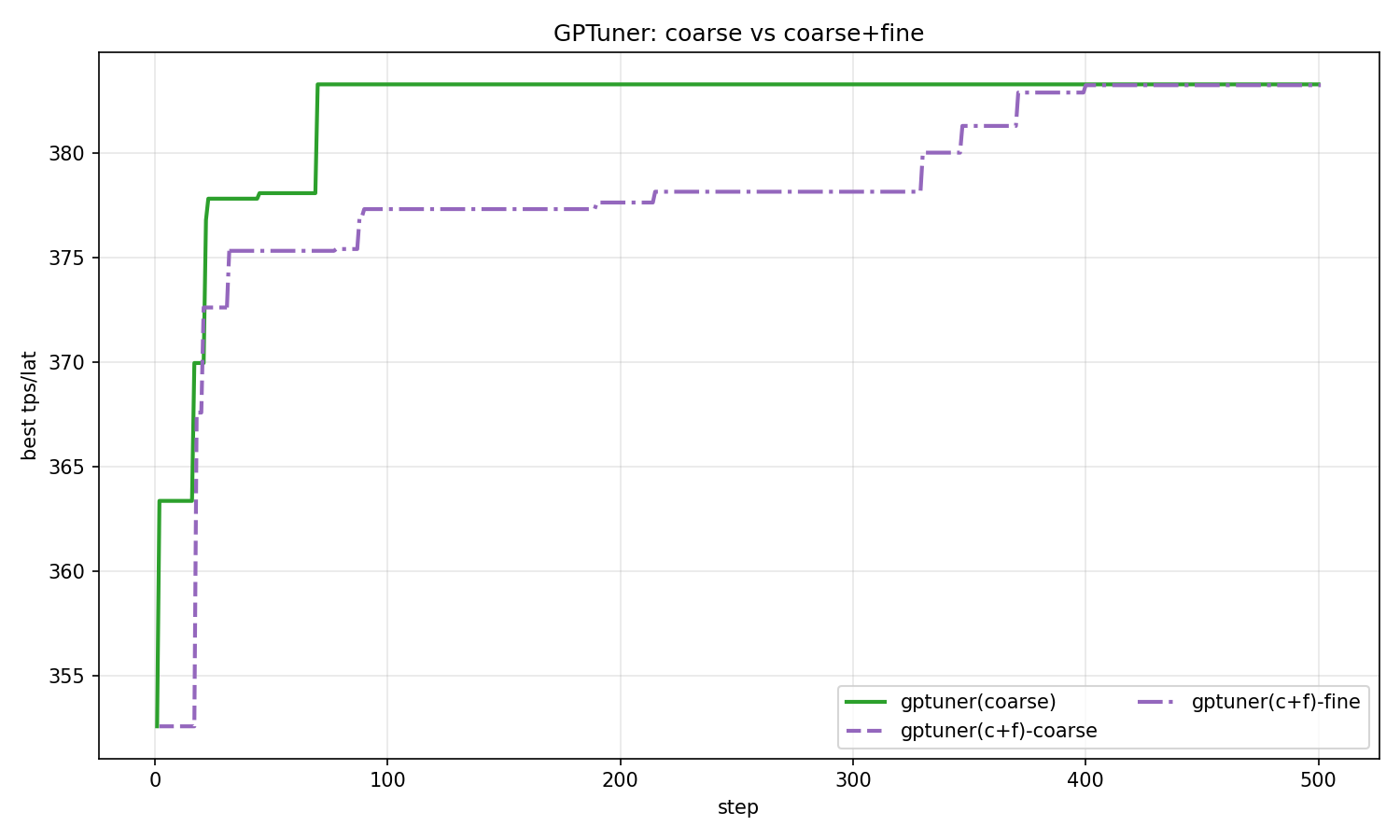}
  \caption{GPTuner:
  coarse-only vs.\ coarse+fine.}
  \label{fig:gpt-coarse-ablation}
\end{figure}

\textbf{coarse-only vs.\ coarse+fine.}
To further examine GPTuner’s tuning behavior, we ran an ablation on a
MySQL read-only workload under 12 cores / 64\,GB RAM / 200\,GB disk.
We compared two schedules: (i) coarse+fine with a 30-step coarse
phase (the original uses \textbf{10} steps) followed by an extended fine phase, and
(ii) coarse-only. As shown in
Figure~\ref{fig:gpt-coarse-ablation}, coarse-only attains the same best
$\mathrm{TPS}/\mathrm{p95}$ but earlier (fewer evaluations). A plausible reason is that
fine begins before coarse has settled near a good basin; once the space
is expanded, SMAC expends budget around less promising regions.
Motivated by this observation, we adopt GPTuner–coarse for Stage~2 in the
rest of our experiments.

\subsubsection{DB-BERT}
\begin{figure}[t]
  \centering
  \includegraphics[width=0.5\columnwidth]{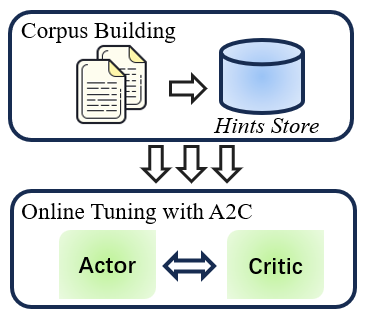} 
  \caption{Overview of dbbert.}
  \label{fig:dbbert}
\end{figure}

\textbf{Sources \& corpus.} DB-BERT constructs a document collection from vendor manuals, engineering blogs, and community forums. For each knob $k$, the system issues targeted queries (e.g., “recommended \texttt{innodb\_buffer\_pool\_size}?”), building a knob-centric evidence set. As illustrated in Figure~\ref{fig:dbbert}, the pipeline centers on corpus building, hint extraction, and online recommendation.

\textbf{Hint extraction \& normalization.} A BERT-style reader (QA) extracts short spans such as “70–80\% of RAM” or “increase to 1000”. A textual classifier normalizes spans into three templates: \emph{absolute} (fixed value), \emph{relative-to-RAM}, and \emph{relative-to-CPU}. The result is a hint table per knob with interpretable formulas.

\textbf{Online hint-guided tuning.} In the online phase, DB-BERT treats each normalized hint for knob $k$ as a base value $v_k$ and lets an A2C controller refine it. At every step the policy selects, for each hinted knob, a \emph{multiplicative factor} $f\in\{0.25,0.5,1,2,4\}$ to scale $v_k$ and an \emph{importance weight} $w\in\{1,2,4,8,16\}$ to prioritize impactful hints; the final configuration is composed from the $f\cdot v_k$ values aggregated by $w$. The system applies the configuration, evaluates the workload, and uses the reward to update the actor–critic. Constraining actions to $(f,w)$ over a small set of hinted knobs enables fast convergence in roughly a few dozen trials.

Given this discrete action design (five multipliers and five weights applied in alternating steps), we experimentally replace A2C with DDPG and TD3 and observed no improvement. The result is that continuous actions must be snapped back to the same five levels, which nullifies the advantage of continuous control and makes the learning signal nearly piecewise constant. Therefore, we retain the original method of dbbert.

\subsection{Stage 3: Dimensionality Reduction and TD3 Fine-Tuning}

On top of the LHS warm start and the LLM-guided recommendations, we add a third stage for RL fine-tuning. Using the uniform samples collected in Stage 1 (the shared ((S,A,P)) pool), we compute knob importances with RF and select the top-(K) knobs; we also fit PCA on the 63-dim state using the same Stage-1 traces to obtain a compact state representation. TD3 then operates on this reduced action/state space to refine the configuration.

\subsubsection{Dimensionality Reduction: RF + PCA}

Following Hunter's dimension reduction method, here we apply two complementary techniques:

\paragraph{ \textbf{RF knob selection}}
MySQL 8.0 exposes \(266\) tunable knobs (e.g., memory: \url{innodb_buffer_pool_size}, \url{key_buffer_size}; concurrency: \url{max_connections}, \url{innodb_thread_concurrency}; I/O: \url{innodb_flush_log_at_trx_commit}, \url{sync_binlog}). Empirically, returns diminish as the number of tuned knobs grows. We therefore perform RF-based feature selection using the \(120\) uniformly distributed LHS samples (Stage~1):
\begin{enumerate}
  \item \textbf{Training data:} \(\mathcal{D}=\{(A^{(j)},P^{(j)})\}_{j=1}^{120}\) with \(A^{(j)}\in\mathbb{R}^{266}\) represents normalized knobs and
  \(p^{(j)}\)is the performance metric based on TPS and p95 latency.
  \item \textbf{RF regression:} We use 100 trees to calculate feature importance, measuring the reduction in mean squared error (MSE) for each knob across the trees.
  \item \textbf{Top-K selection:} Based on feature importance, we rank all 266 knobs and select the top-20 knobs for tuning. This selection strikes an optimal balance between expressiveness and training efficiency, in line with HUNTER’s approach. Tuning more than 20 knobs results in less than 2\% additional gain but significantly higher computational cost.
\end{enumerate}

LHS’s stratified coverage is key: without uniformity, importance estimates would be biased toward over-sampled regions.

\paragraph{\textbf{ PCA state compression}.}
Using the 63 internal metrics (Table~\ref{tab:innodb-metrics}) as-is burdens the critic (curse of dimensionality; correlated metrics such as \url{buffer_pool_reads} and \url{buffer_data_reads}). We apply PCA:
\begin{enumerate}
  \item Standardize each metric to zero mean/unit variance across the 120 LHS samples.
  \item Covariance matrix \(\Sigma\in\mathbb{R}^{63\times63}\), and perform eigendecomposition of  \(\Sigma\).
  \item Select top-13 components explaining \(\sim95\%\) variance (as in Hunter).
\end{enumerate}
This reduces critic input from \(63+266=329\) to \({13+20=33}\), yielding faster, stabler learning. The transforms are computed once from Stage~1 data and reused in TD3.

\subsubsection{TD3 Algorithm Overview}
TD3 is off-policy reinforcement learning algorithm designed for continuous action spaces. It builds upon DDPG (Deep Deterministic Policy Gradient) by introducing three key mechanisms to improve training stability and reduce overestimation bias.

\begin{figure}[t]
  \centering
  \includegraphics[width=\columnwidth]{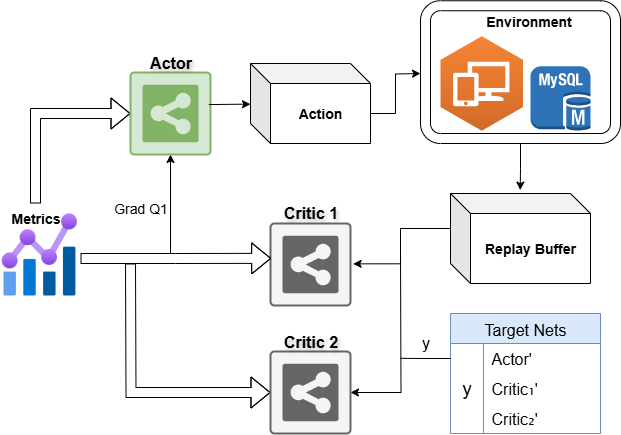} 
  \caption{the struction of TD3 model}
  \label{fig:td32}
\end{figure}

As shown in Figure~\ref{fig:td32}, the TD3 algorithm consists of three key components:

\begin{itemize}
\item \textbf{Actor Network:} The actor takes the current state of the system, represented by various metrics such as TPS, latency, and other database performance indicators, and outputs an action— a set of configuration values for the database knobs. This action guides the tuning process by adjusting the database's parameters in response to observed performance. The actor is updated continuously by the gradients computed from the critic's evaluations.

\item \textbf{Critic Networks:} Critic 1 and Critic 2 are two independent Q-value estimators that assess the quality of the actions selected by the actor. They take the current state and the selected action as input, outputting Q-values that estimate the expected future reward for those state-action pairs. By using two critics, TD3 reduces overestimation bias—a common issue in Q-learning, where overestimation of action values can destabilize training. The target for both critics is updated using the minimum of the two Q-values, providing a more conservative estimate.

\item \textbf{Replay Buffer and Target Networks:} The replay buffer stores past experiences of the system—state-action-reward transitions which are used to train the networks. The target networks are slowly updated copies of the main networks. This gradual update helps stabilize training by preventing rapid changes in the target values, which would otherwise cause oscillations or instability in the learning process.
\end{itemize}

\subsubsection{Empirical Validation: DDPG vs.\ TD3}
\begin{figure}[t]
  \centering
  \includegraphics[width=\columnwidth]{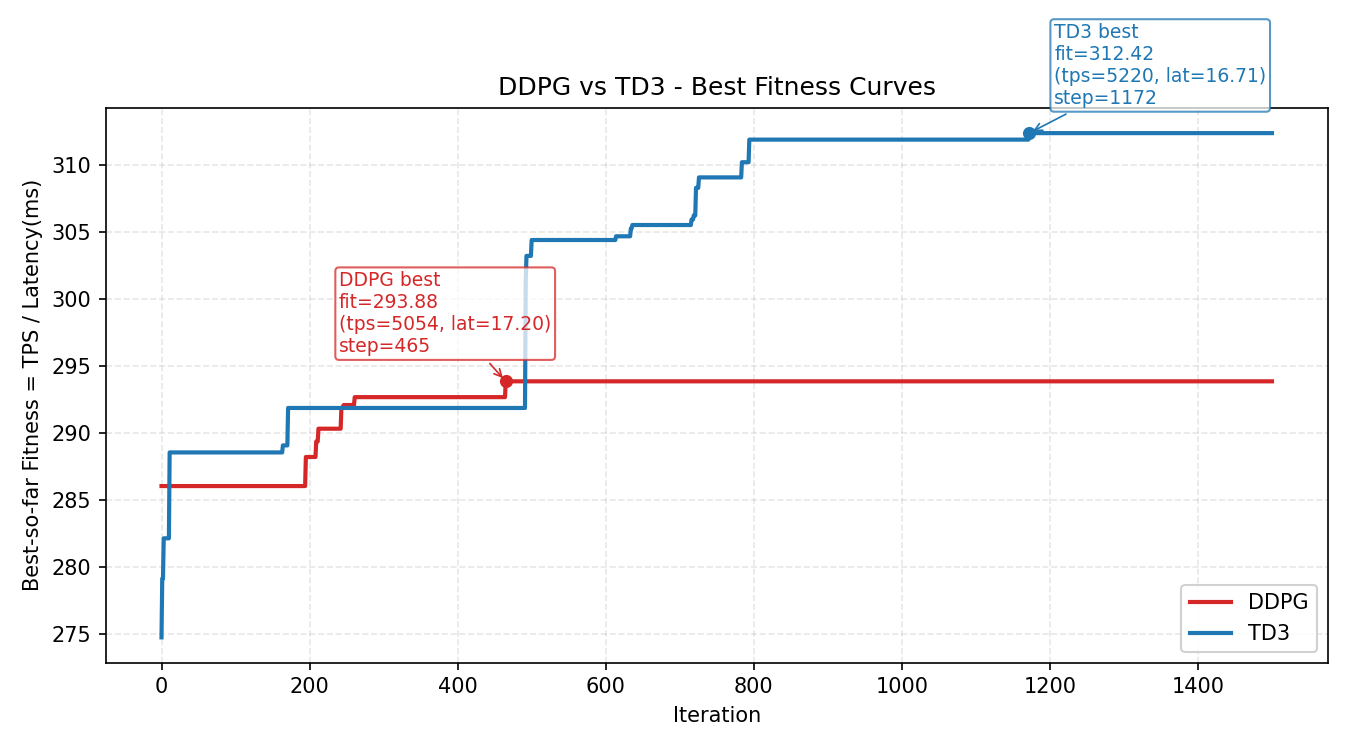} 
  \caption{the struction of TD3 model}
  \label{fig:ddpg_vs_td3}
\end{figure}

To empirically validate TD3’s superiority over DDPG for database tuning, we conducted a controlled comparison on the sysbench read workload, starting from the same LHS warm-start initialization and deliberately omitting any LLM-guided Stage-2 recommendations. Both algorithms were trained with identical hyperparameters (learning rates, batch size, replay buffer capacity).
Figure~\ref{fig:ddpg_vs_td3} shows the best fitness curves over 1500 training iterations.

As illustrated in Figure~\ref{fig:ddpg_vs_td3}, DDPG achieves a best fitness of \textbf{293.88} (TPS \(=5054\), latency \(=17.20\)\,ms) at step 465, while TD3 reaches \textbf{312.42} (TPS \(=5220\), latency \(=16.71\)\,ms) at step 1172, representing a \textbf{6.3\%} improvement in overall performance.

The performance boost comes from TD3's improvements, including twin critic networks, which reduce overestimation bias and lead to more accurate Q-value estimates. Target policy smoothing stabilizes learning by preventing the actor from exploiting narrow peaks in the Q-function. Finally, delayed updates allow the critic networks to stabilize before the actor is updated, making TD3 more efficient and stable compared to DDPG, which updates both networks simultaneously.

\subsubsection{Pipeline Summary}

L2T-Tune integrates three stages into a unified tuning pipeline:

\begin{enumerate}
  \item \textbf{Stage 1 (LHS Warm-Start)}: Generate 120 uniformly distributed configurations via Latin Hypercube Sampling, establishing a baseline and collecting diverse \((S,A,P)\) samples for later dimensionality reduction.
    \item \textbf{Stage 2 (LLM-Guided Optimization):} Starting from the best LHS configuration, apply LLM model documentation-driven recommendations (vendor manuals/ blogs/ forums) to rapidly improve the baseline with few trials.
  \item \textbf{Stage 3 (TD3 Fine-Tuning)}: Use the 120 LHS samples to perform Random-Forest knob selection and PCA state compression; then fine-tune with TD3 from the Stage-2 configuration.
\end{enumerate}

This three-stage design is compositional: LHS provides uniform exploration and training data, the LLM stage delivers fast convergence from textual expertise, and TD3 performs fine-grained optimization in the reduced space.

\section{Experiments}

\paragraph{\textbf{Experiment Setting}}
The experiments ran on a single x86\_64 KVM virtual machine (Ubuntu 24.04) provisioned with 12 vCPUs on an AMD EPYC 7713 host, 64 GB RAM, and a 120 GB SSD; no GPU accelerators were used. All experiments were run on this single machine setup, ensuring consistency across different test cases and providing a standard environment for evaluating the performance of the tuning algorithms.

We conducted experimental comparisons with state-of-the-art tuners from both RL-based and LLM-based lines, including CDBTune, HUNTER, DB-BERT, and GPTuner, as well as the coarse-only GPTuner variant (GPTuner–coarse) discussed above.

\begin{itemize}
  \item \textbf{CDBTune:} CDBTune utilizes deep reinforcement learning to adaptively tune database configurations, providing efficient optimization.
  \item \textbf{HUNTER:} HUNTER combines genetic algorithms with deep reinforcement learning to perform hybrid tuning, offering robust performance across varying database configurations.
  \item \textbf{DB-BERT:} DB-BERT leverages a large language model to extract tuning hints from documentation, significantly improving performance with fewer trials.
  \item \textbf{GPTuner:} GPTuner integrates GPT-based models with Bayesian optimization to guide database tuning, offering fast convergence and enhanced tuning accuracy.
\end{itemize}

These models were selected for comparison to evaluate the effectiveness of our proposed approach in various database tuning scenarios.

\paragraph{\textbf{Performance Comparison}}
We first conducted experiments on the MySQL database under four different workloads: Sysbench read-only, Sysbench read-write (1:1), and Sysbench write-only, as well as a TPC-C read-write workload. Each experiment was performed with a 60-second stress test. The database setup consisted of 32 tables, each containing 100,000 rows. For Sysbench read-only, the test was run with 64 threads, while the other workloads used 32 threads. The evaluation metric for all experiments was $\mathrm{TPS}/\mathrm{p95}$. These experiments were designed to evaluate the performance of the tuning models under various database load conditions.

\begin{figure*}[t]
  \centering
  \includegraphics[width=\textwidth]{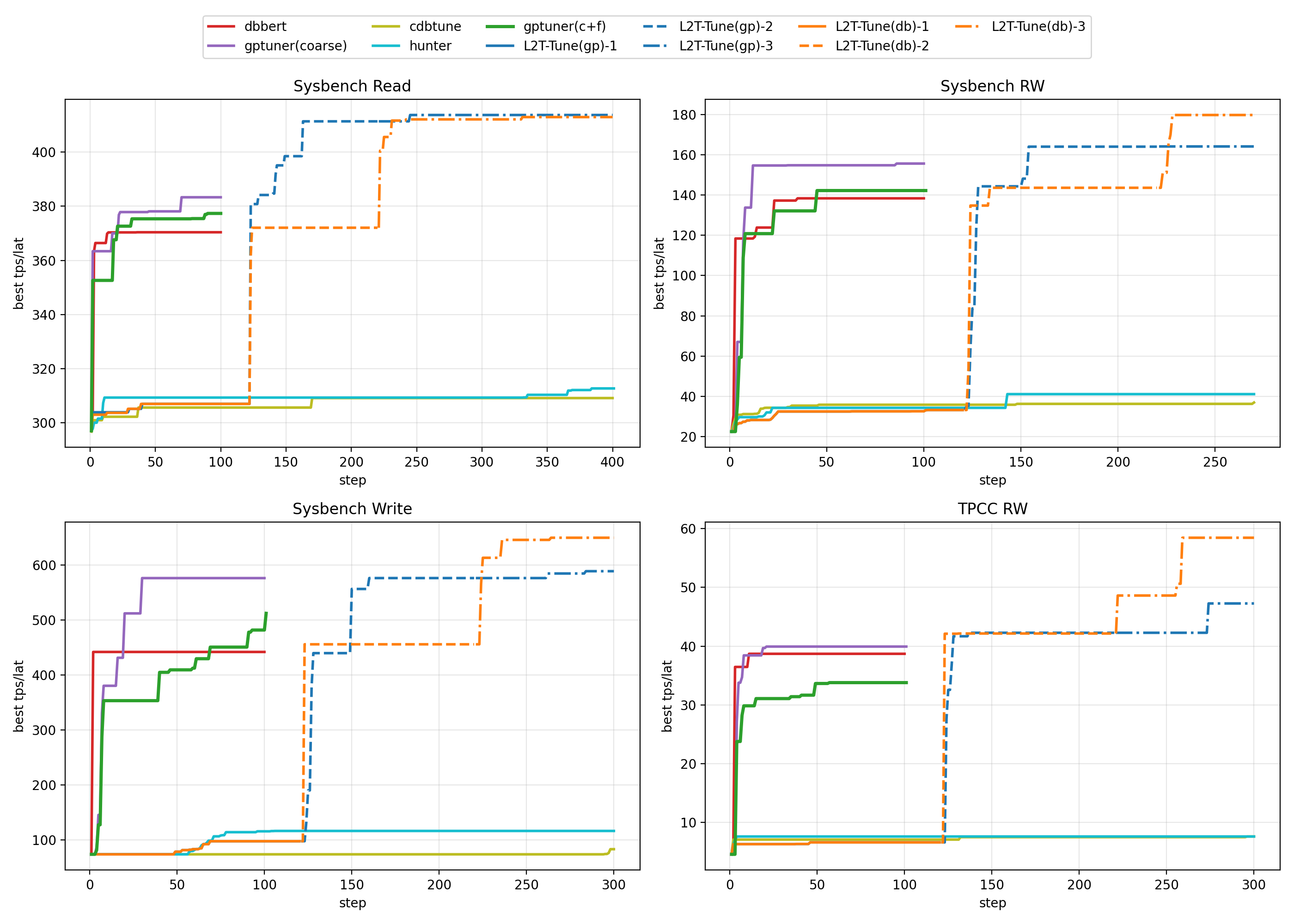} 
  \caption{Performance comparison of different tuning models across four workloads: Sysbench read-only, Sysbench read-write (1:1), Sysbench write-only, and TPC-C read.}
  \label{fig:result}
\end{figure*}

\begin{table}[htbp]
\centering
\caption{Performance Summary of Different Models Across Workloads}
\label{tab:performance_summary}
\begin{tabular}{|l|l|l|l|l|}
\hline
\textbf{Workload} & \textbf{Model} & \textbf{Best TPS/p95} & \textbf{Steps} \\
\hline
 &  \textbf{L2T-Tuner(db)} &  \textbf{412.92 }&  \textbf{331} \\
 &  \textbf{L2T-Tuner(gp)} &  \textbf{413.68 }&  \textbf{245} \\
&  GPTuner(c+f) & 377.34 & 90 \\
Sysbench Read &  GPTuner(c) & 383.30 & 70 \\
& DBBert & 370.38 & 36 \\
 & Hunter & 312.11 & 368 \\
&  CDBTune & 309.14 & 170 \\
\hline
 &   \textbf{L2T-Tuner(db)} &  \textbf{179.69} &  \textbf{228} \\
  &   \textbf{L2T-Tuner(gp)} &  \textbf{163.98} &  \textbf{154} \\
& GPTuner(c+f) & 142.19 & 45 \\
Sysbench RW & GPTuner(c) & 155.59 & 86 \\
&  DBBert & 138.33 & 35 \\
&  Hunter & 41.13 & 143 \\
 &  CDBTune & 36.96 & 270 \\
\hline
 & \textbf{L2T-Tuner(db)} & \textbf{649.72} & \textbf{264} \\
 & \textbf{L2T-Tuner(gp)} & \textbf{588.83} & \textbf{284} \\
 & GPTuner(c+f) & 514.57 & 100 \\
Sysbench Write& GPTuner(c) & 576.24 & 30 \\
 & DBBert & 441.85 & 2 \\
& Hunter & 116.06 & 106 \\
 &  CDBTune & 82.86 & 298 \\

\hline
&  \textbf{L2T-Tuner(db)} &\textbf{58.45} & \textbf{259} \\
&  \textbf{L2T-Tuner(gp)} &\textbf{47.26} & \textbf{274} \\
 &  GPTuner(c+f) & 33.78 & 54 \\
 TPC-C RW &  GPTuner(c) & 39.92 & 21 \\
 & DBBert & 38.69 & 11 \\
&  Hunter & 7.59 & 4 \\
&  CDBTune & 7.60 & 296 \\
\hline
\end{tabular}
\end{table}

The results in Figure~\ref{fig:result} show a clear separation among methods. Both variants of L2T-Tune lead overall; L2T-Tune (db) attains the highest final fitness, with L2T-Tune (gp) second. To probe the ceiling of Stage~2, we allotted 100 steps to the LLM stage in both variants. Empirically, the LLM stage of L2T-Tune (db) peaks within $\leq 20$ steps (often $\leq 5$) because its resource-aware text prior maps relative rules (e.g., ``25\% RAM'') to machine-specific values and narrows the candidate set to fewer but more accurate knobs. By contrast, L2T-Tune (gp) explores a broader set of concrete knobs and typically needs $\sim 50$ steps to surpass its early baseline, even though its Stage~2 peak can sometimes exceed the db variant. In Stage~3, a TD3 agent performs critic-guided fine-tuning over the LLM-guided basin, which quickly closes any residual gap for L2T-Tune (db) ($\approx 30$ steps) and amplifies its lead; L2T-Tune (gp) also converges in $\sim 30$ Stage~3 steps but gains less. 

Importantly, \emph{coarse-only (c)} often reaches a strong configuration faster and sometimes higher than \emph{coarse+fine (c+f)}. The reason is that the coarse budget is too small; consequently, fine optimization starts before coarse lands near a good basin, leading to premature, locally biased refinement. In L2T-Tune we deliberately use \emph{coarse} as the Stage~2 recommender and delegate fine-grained adjustment to Stage~3 \emph{TD3}; coupled with Stage~1 warm-start that filters knobs by hardware-aware priors, the RL agent operates in a compact, high-signal subspace and thus surpasses GPTuner in both speed and final quality.

Quantitatively (Table~\ref{tab:performance_summary}), relative to the strongest baseline GPTuner (coarse+fine), our best L2T-Tune variant improves TPS/p95 by +22.6\% on \textbf{Sysbench Read} (L2T-Tune(gp) 413.68 vs 377.34), +26.4\% on \textbf{Sysbench RW} (L2T-Tune(db) 179.69 vs 142.19), +26.3\% on \textbf{Sysbench Write} (L2T-Tune(db) 649.72 vs 514.56), and +73.0\% on \textbf{TPC-C RW} (L2T-Tune(db) 58.45 vs 33.78), averaging a \textbf{+37.1\%} uplift. Compared with HUNTER and CDBTune, inserting the LLM stage dramatically accelerates RL convergence; within the same budget on a single server, RL-only systems struggle to reach competitive levels.

Step budgets that achieve the best results in our runs are:
\begin{itemize}
   
 \item \textbf{L2T-Tune(db):} 120 (LHS) + 5 (LLM) + 30 (TD3) = 155 steps.
 \item \textbf{L2T-Tune(gp):} 120 (LHS) + 50 (LLM) + 30 (TD3) = 200 steps.

\end{itemize}

\begin{figure}[t]
  \centering
  \includegraphics[width=\columnwidth]{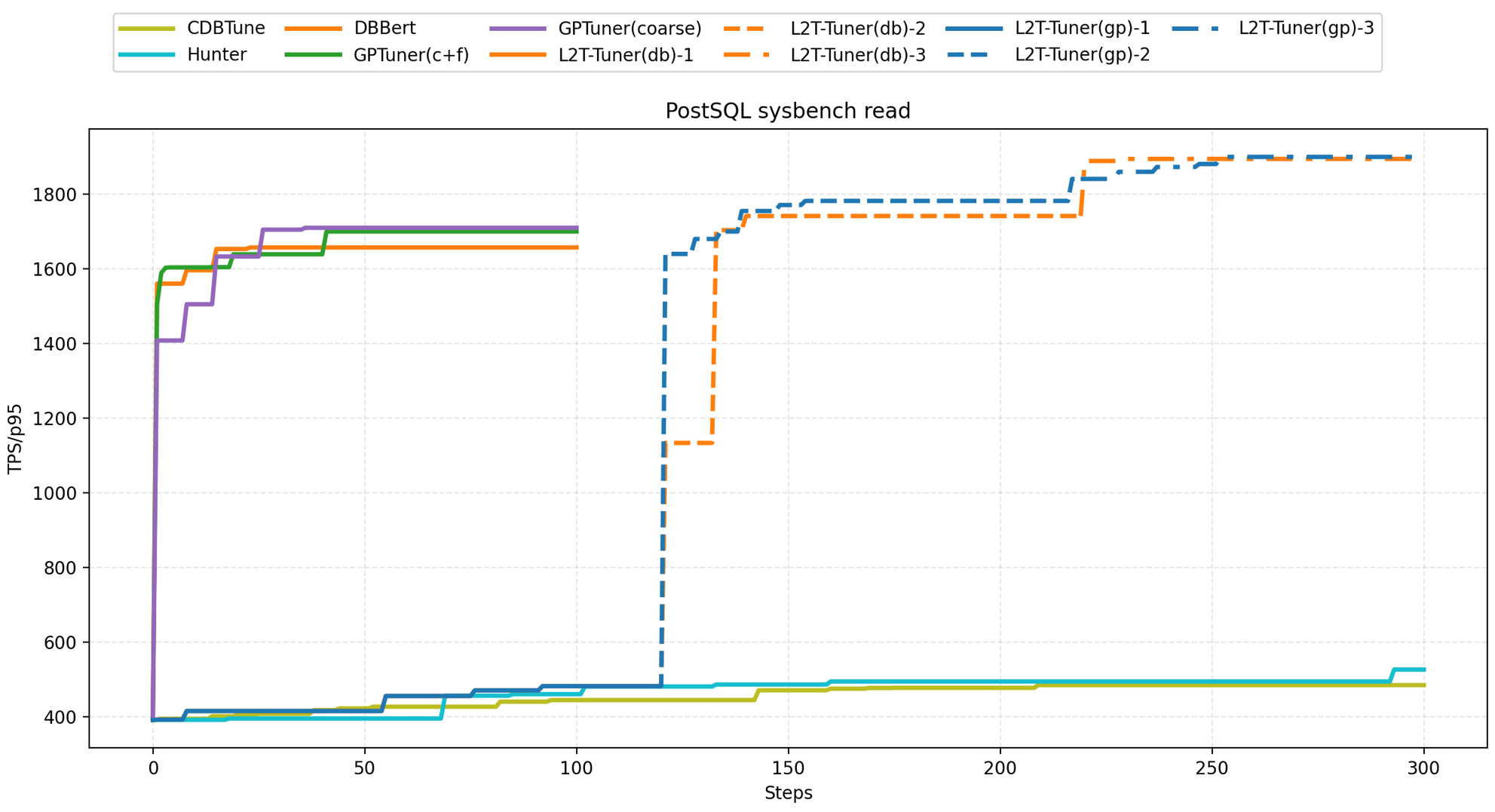} 
  \caption{Sysbench read experiment on PostgreSQL.}
  \label{fig:post}
\end{figure}

Additionally, we conducted similar experiments using PostgreSQL with Sysbench read-only workload. As shown in Figure~\ref{fig:post}, the model performance followed the same trend as observed in the MySQL experiments.

\paragraph{\textbf{Adaptability}}

In both the cdbtune and hunter frameworks, extensive research has demonstrated that their respective models exhibit remarkable transferability during the online tuning phase. Nevertheless, achieving optimal performance often necessitates several hours of online tuning training.

In response to this challenge, we have developed a novel semi-transfer approach based on our model. This method involves transferring only the hot-start component and subsequently leveraging dbbert to effect rapid improvement over a relatively short period. 

\begin{figure*}[t]
  \centering
  \includegraphics[width=\textwidth]{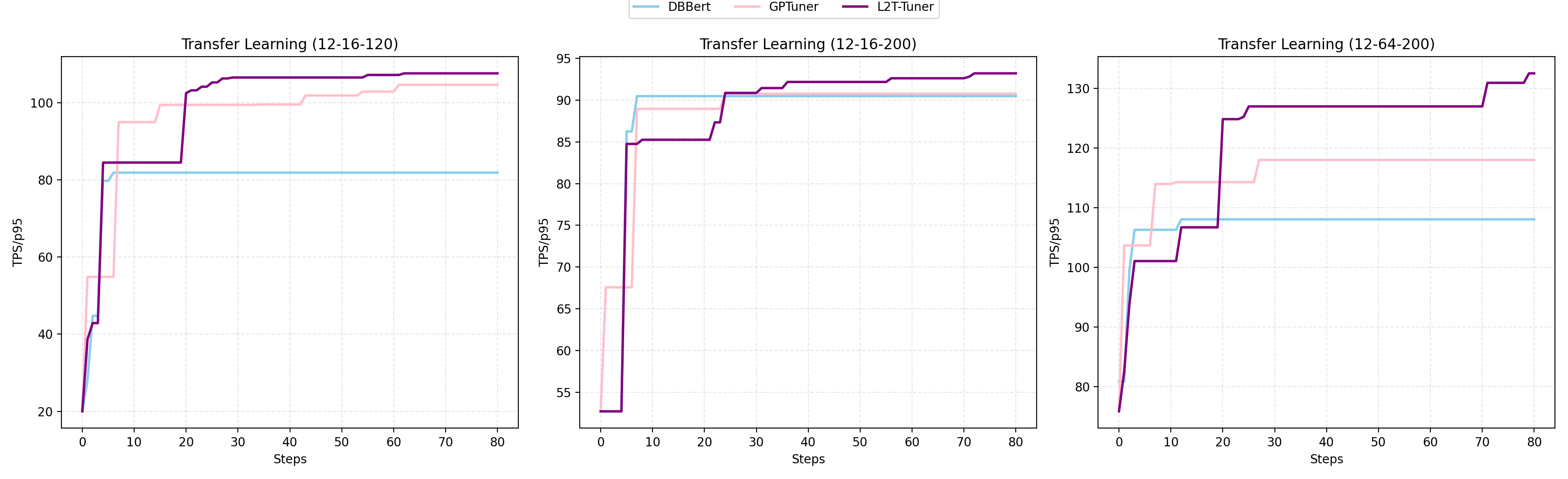} 
  \caption{Transfer learning performance comparison using the semi-transfer approach.}
  \label{fig:trans3}
\end{figure*}

To evaluate the effectiveness of our proposed method, we conducted a series of experiments. As shown in Figure~\ref{fig:trans3}, we adjusted the sysbench read data across different configurations, specifically moving from a setup of 12 cores, 64g memory, and 200g disk to conditions of 12 cores, 16g memory, 120g disk, as well as comparing with the 12 - 16 - 200 and 12 - 64 - 200 setups. These configurations were benchmarked against dbbert and gptuner.

The results of our study indicate that our model can essentially reach the optimal value within 30 steps. This is achieved through 15 steps of dbbert operation followed by 15 steps of TD3 tuning. Moreover, the performance of our model outperforms that of dbbert and gptuner, highlighting the superiority of our proposed semi - transfer method in the context of online tuning.

\section{Related work}
Modern DBMS performance relies heavily on configuration tuning~\cite{chaudhuri1997efficient,ding2020alex,galakatos2019fiting,kraska2018case,liu2020distributed,ma2018query,ma2021mb2,sadri2020online,schnaitter2010semi,tan2019ibtune,van2017automatic,van2021inquiry}, as production systems expose hundreds of interdependent "knobs" that impact key metrics like throughput (TPS), tail latency (p95), and query rate (QPS). Selecting optimal settings is inherently difficult, as tuning involves high-dimensional, combinatorial (NP-hard) optimization with complex, non-transparent interactions between parameters. The tuning process is essentially a black-box, where small changes to the knobs can significantly affect performance.

In practice, DBAs typically optimize for TPS and p95, but even minor adjustments can lead to dramatic shifts. Recent work has explored various strategies, including search-based methods, supervised learning, reinforcement learning (with offline and online adaptation), and LLM-guided recommendations. In the following sections, we review these approaches and present our three-stage method.

\textbf{ Machine Learning–Based Methods:}
OtterTune\cite{van2017automatic} models the mapping from database state to configuration with Gaussian Processes, using internal metrics (e.g., cache utilization, read/write activity) to predict knob settings rather than relying only on TPS/latency.

\textbf{RL Based Methods:}
CDBTune \cite{CDBTune} frames tuning as continuous control and trains a DDPG agent with offline data, followed by online fine-tuning to adapt the learned policy to new workloads and hardware.
HUNTER \cite{HUNTER} augments RL with a GA-style preheating search and dimension reduction (RF/PCA), then runs population-based RL to refine the reduced set of impactful knobs.

\textbf{LLM Driven Approaches:} LLM Driven Approaches
DB-BERT \cite{DB-BERT} builds a corpus of manuals/blogs/forums, extracts text spans via QA, and converts them into knob recommendations that serve as high-quality hints.
GPTuner \cite{GPTuner} prompts GPT with workload/state summaries and couples the LLM’s suggestions with Bayesian optimization to select concrete knob values.
LLMTune \cite{LLMTune} uses transformer-based reading of vendor docs and community posts to map natural-language guidance to actionable configuration updates.
$\lambda$-Tune \cite{lambdatune} / LaTuner \cite{Latuner} employ LLM reasoning to generate rule-like knob policies and deploy them as lightweight cloud functions for rapid online adjustments.
E2ETune \cite{huang2024e2etune} provides an end-to-end automated pipeline that integrates LLM-derived guidance with learning-based controllers to update knobs continuously during execution.

\textbf{Hybrid Approaches:}
L2T-Tune (ours) follows the CDBTune/HUNTER offline-training + online fine-tuning paradigm but couples it with LLM guidance: Stage 1 uses LHS for a uniform warm start; Stage 2 applies LLM models for fast, documentation-driven recommendations; Stage 3 reduces dimensionality (RF/PCA) and fine-tunes with TD3, yielding stronger optimization within the same framework.

\section{Conclusion}

In this paper, we presented L2T-Tune, a three-stage hybrid tuner that couples uniform LHS warm-start, LLM-guided recommendation (GPTuner or DB-BERT), and TD3 fine-tuning on RF/PCA-reduced spaces. This design turns expensive blind exploration into a short, knowledge-steered search and then applies stable continuous control to polish the result.

Empirically, across four MySQL workloads (and a PostgreSQL read case), L2T-Tune delivers state-of-the-art TPS/p95: it achieves up to \textbf{+73.0\%} improvement over the best baseline and averages \textbf{+37.1\%} across all workloads; it converges on a single server without large parallel fleets, and in semi-transfer online tuning reaches high-quality configurations in $\approx 30$ evaluations.

L2T-Tune is closest to the CDBTune/HUNTER lineage but substantially enhances that framework’s efficiency and final quality by inserting an LLM stage and by narrowing RL to the most impactful features and knobs. The result is a practical pipeline that tunes well from scratch, adapts quickly under hardware changes (semi-transfer), and keeps resource costs modest.

\section*{Acknowledgment}

The work was supported by Strategy Priority Research Program (Supported by the Strategic
Priority Research Program of the Chinese Academy of Sciences) under Grant No.XDA0360202.


\begin{thebibliography}{00}


\bibitem{CDBTune} J. Zhang, Y. Liu, K. Zhou, G. Li, Z. Xiao, B. Cheng, J. Xing, Y. Wang, T. Cheng, L. Liu, et al., ``An end-to-end automatic cloud database tuning system using deep reinforcement learning,'' in Proceedings of the 2019 international conference on management of data, 2019, pp. 415--432.

\bibitem{duan2009tuning} S. Duan, V. Thummala, and S. Babu, ``Tuning database configuration parameters with ituned,'' Proceedings of the VLDB Endowment, vol. 2, no. 1, pp. 1246--1257, 2009.

\bibitem{BestConfig} Y. Zhu, J. Liu, M. Guo, Y. Bao, W. Ma, Z. Liu, K. Song, and Y. Yang, ``Bestconfig: tapping the performance potential of systems via automatic configuration tuning,'' in Proceedings of the 2017 symposium on cloud computing, 2017, pp. 338--350.


\bibitem{van2017automatic} D. Van Aken, A. Pavlo, G. J. Gordon, and B. Zhang, ``Automatic database management system tuning through large-scale machine learning,'' in Proceedings of the 2017 ACM international conference on management of data, 2017, pp. 1009--1024.

\bibitem{gallinucci2019sparktune} E. Gallinucci and M. Golfarelli, ``SparkTune: Tuning spark SQL through query cost modeling,'' in Advances in Database Technology-EDBT 2019, 22th International Conference on Extending Database Technology, Proceedings, 2019, pp. 546--549.

\bibitem{cgptuner} S. Cereda, S. Valladares, P. Cremonesi, and S. Doni, ``Cgptuner: a contextual gaussian process bandit approach for the automatic tuning of it configurations under varying workload conditions,'' Proceedings of the VLDB Endowment, vol. 14, no. 8, pp. 1401--1413, 2021.

\bibitem{zhang2021restune} X. Zhang, H. Wu, Z. Chang, S. Jin, J. Tan, F. Li, T. Zhang, and B. Cui, ``Restune: Resource oriented tuning boosted by meta-learning for cloud databases,'' in Proceedings of the 2021 international conference on management of data, 2021, pp. 2102--2114.

\bibitem{zhang2022towards} X. Zhang, H. Wu, Y. Li, J. Tan, F. Li, and B. Cui, ``Towards dynamic and safe configuration tuning for cloud databases,'' in Proceedings of the 2022 International Conference on Management of Data, 2022, pp. 631--645.




\bibitem{HUNTER} B. Cai, Y. Liu, C. Zhang, G. Zhang, K. Zhou, L. Liu, C. Li, B. Cheng, J. Yang, and J. Xing, ``HUNTER: an online cloud database hybrid tuning system for personalized requirements,'' in Proceedings of the 2022 International Conference on Management of Data, 2022, pp. 646--659.

\bibitem{li2019qtune} G. Li, X. Zhou, S. Li, and B. Gao, ``Qtune: A query-aware database tuning system with deep reinforcement learning,'' Proceedings of the VLDB Endowment, vol. 12, no. 12, pp. 2118--2130, 2019.






\bibitem{lillicrap2015continuous} T. P. Lillicrap, J. J. Hunt, A. Pritzel, N. Heess, T. Erez, Y. Tassa, D. Silver, and D. Wierstra, ``Continuous control with deep reinforcement learning,'' arXiv preprint arXiv:1509.02971, 2015.

\bibitem{watkins1992q} C. J. C. H. Watkins and P. Dayan, ``Q-learning,'' Machine learning, vol. 8, no. 3, pp. 279--292, 1992.

\bibitem{van2016deep} H. Van Hasselt, A. Guez, and D. Silver, ``Deep reinforcement learning with double q-learning,'' in Proceedings of the AAAI conference on artificial intelligence, vol. 30, no. 1, 2016.

\bibitem{kanellis2020too} K. Kanellis, R. Alagappan, and S. Venkataraman, ``Too many knobs to tune? towards faster database tuning by pre-selecting important knobs,'' in 12th USENIX Workshop on Hot Topics in Storage and File Systems (HotStorage 20), 2020.

\bibitem{DB-BERT} I. Trummer, ``DB-BERT: a Database Tuning Tool that Reads the Manual,'' in Proceedings of the 2022 international conference on management of data, 2022, pp. 190--203.

\bibitem{LLMTune} X. Huang, H. Li, J. Zhang, X. Zhao, Z. Yao, Y. Li, Z. Yu, T. Zhang, H. Chen, and C. Li, ``Llmtune: Accelerate database knob tuning with large language models,'' CoRR, 2024.

\bibitem{GPTuner} J. Lao, Y. Wang, Y. Li, J. Wang, Y. Zhang, Z. Cheng, W. Chen, M. Tang, and J. Wang, ``GPTuner: An LLM-Based Database Tuning System,'' ACM SIGMOD Record, vol. 54, no. 1, pp. 101--110, 2025.

\bibitem{mckay2000comparison} M. D. McKay, R. J. Beckman, and W. J. Conover, ``A comparison of three methods for selecting values of input variables in the analysis of output from a computer code,'' Technometrics, vol. 42, no. 1, pp. 55--61, 2000.




\bibitem{fujimoto2018addressing} S. Fujimoto, H. Hoof, and D. Meger, ``Addressing function approximation error in actor-critic methods,'' in International conference on machine learning, 2018, pp. 1587--1596.


\bibitem{lambdatune} V. Giannakouris and I. Trummer, ``$\lambda$-tune: Harnessing large language models for automated database system tuning,'' Proceedings of the ACM on Management of Data, vol. 3, no. 1, pp. 1--26, 2025.



\bibitem{RF} L. Breiman, ``Random forests,'' Machine learning, vol. 45, no. 1, pp. 5--32, 2001.


\bibitem{PCA} G. H. Dunteman, Principal components analysis, vol. 69. Sage, 1989.




\bibitem{chaudhuri1997efficient} S. Chaudhuri and V. R. Narasayya, ``An efficient, cost-driven index selection tool for Microsoft SQL server,'' in VLDB, vol. 97, 1997, pp. 146--155.

\bibitem{ding2020alex} J. Ding, U. F. Minhas, J. Yu, C. Wang, J. Do, Y. Li, H. Zhang, B. Chandramouli, J. Gehrke, D. Kossmann, et al., ``ALEX: an updatable adaptive learned index,'' in Proceedings of the 2020 ACM SIGMOD international conference on management of data, 2020, pp. 969--984.

\bibitem{galakatos2019fiting} A. Galakatos, M. Markovitch, C. Binnig, R. Fonseca, and T. Kraska, ``Fiting-tree: A data-aware index structure,'' in Proceedings of the 2019 international conference on management of data, 2019, pp. 1189--1206.

\bibitem{kraska2018case} T. Kraska, A. Beutel, E. H. Chi, J. Dean, and N. Polyzotis, ``The case for learned index structures,'' in Proceedings of the 2018 international conference on management of data, 2018, pp. 489--504.

\bibitem{liu2020distributed} J. Liu and C. Zhang, ``Distributed learning systems with first-order methods,'' Foundations and Trends in Databases, vol. 9, no. 1, pp. 1--100, 2020.

\bibitem{ma2018query} L. Ma, D. Van Aken, A. Hefny, G. Mezerhane, A. Pavlo, and G. J. Gordon, ``Query-based workload forecasting for self-driving database management systems,'' in Proceedings of the 2018 International Conference on Management of Data, 2018, pp. 631--645.

\bibitem{ma2021mb2} L. Ma, W. Zhang, J. Jiao, W. Wang, M. Butrovich, W. S. Lim, P. Menon, and A. Pavlo, ``MB2: decomposed behavior modeling for self-driving database management systems,'' in Proceedings of the 2021 International Conference on Management of Data, 2021, pp. 1248--1261.

\bibitem{sadri2020online} Z. Sadri, L. Gruenwald, and E. Leal, ``Online index selection using deep reinforcement learning for a cluster database,'' in 2020 IEEE 36th International Conference on Data Engineering Workshops (ICDEW), 2020, pp. 158--161.

\bibitem{schnaitter2010semi} K. Schnaitter and N. Polyzotis, ``Semi-automatic index tuning: Keeping dbas in the loop,'' arXiv preprint arXiv:1004.1249, 2010.

\bibitem{tan2019ibtune} J. Tan, T. Zhang, F. Li, J. Chen, Q. Zheng, P. Zhang, H. Qiao, Y. Shi, W. Cao, and R. Zhang, ``ibtune: Individualized buffer tuning for large-scale cloud databases,'' Proceedings of the VLDB Endowment, vol. 12, no. 10, pp. 1221--1234, 2019.

\bibitem{van2021inquiry} D. Van Aken, D. Yang, S. Brillard, A. Fiorino, B. Zhang, C. Bilien, and A. Pavlo, ``An inquiry into machine learning-based automatic configuration tuning services on real-world database management systems,'' Proceedings of the VLDB Endowment, vol. 14, no. 7, pp. 1241--1253, 2021.



\bibitem{Latuner} C. Fan, Z. Pan, W. Sun, C. Yang, and W.-N. Chen, ``Latuner: An llm-enhanced database tuning system based on adaptive surrogate model,'' in Joint European Conference on Machine Learning and Knowledge Discovery in Databases, 2024, pp. 372--388.

\bibitem{huang2024e2etune} X. Huang, H. Li, J. Zhang, X. Zhao, Z. Yao, Y. Li, T. Zhang, J. Chen, H. Chen, and C. Li, ``E2etune: End-to-end knob tuning via fine-tuned generative language model,'' arXiv preprint arXiv:2404.11581, 2024.

\bibitem{lindauer2022smac3} M. Lindauer, K. Eggensperger, M. Feurer, A. Biedenkapp, D. Deng, C. Benjamins, T. Ruhkopf, R. Sass, and F. Hutter, ``SMAC3: A versatile Bayesian optimization package for hyperparameter optimization,'' Journal of Machine Learning Research, vol. 23, no. 54, pp. 1--9, 2022.







\bibitem{lu2019speedup} J. Lu, Y. Chen, H. Herodotou, and S. Babu, ``Speedup your analytics: Automatic parameter tuning for databases and big data systems,'' Proceedings of the VLDB Endowment, vol. 12, no. 12, pp. 1970--1973, 2019.

\bibitem{li2019xuanyuan} G. Li, X. Zhou, and S. Li, ``Xuanyuan: An ai-native database,'' IEEE Data Eng. Bull., vol. 42, no. 2, pp. 70--81, 2019.

\bibitem{chen2019data} J. Chen, Y. Chen, Z. Chen, A. Ghazal, G. Li, S. Li, W. Ou, Y. Sun, M. Zhang, and M. Zhou, ``Data management at huawei: Recent accomplishments and future challenges,'' in 2019 IEEE 35th International Conference on Data Engineering (ICDE), 2019, pp. 13--24.





\end{thebibliography}
\end{document}